\documentclass[10pt]{article}
\usepackage{amsmath}
\usepackage{mathrsfs}
\usepackage{braket}
\usepackage{ascmac}
\usepackage{bm}
\usepackage[dvipdfmx]{graphicx}
\usepackage[at]{easylist}
\usepackage{here}
\usepackage{cite}
\usepackage{amssymb}
\usepackage{mhchem}
\usepackage{physics}
\usepackage{authblk}
\usepackage{indentfirst}
\usepackage[subrefformat=parens]{subcaption}
\usepackage[top=30truemm,bottom=30truemm,left=20truemm,right=20truemm]{geometry}
\usepackage{color}
\makeatletter
\@addtoreset{equation}{section}

\makeatother
\definecolor{DarkBlue}{rgb}{0,0,0.9} 

\definecolor{DarkRed}{rgb}{0.65,0,0}

\title{Stationary Vacuum Bubble in a Kerr-de Sitter Spacetime}
\date{ }
\author{Daiki Saito\footnote{Email:saito.daiki.g3@s.mail.nagoya-u.ac.jp}~}
\author{Chul-Moon Yoo\footnote{Email:yoo.chulmoon.k6@f.mail.nagoya-u.ac.jp}}
\affil{Division of Science, Graduate School of Science, Nagoya University, Nagoya 464-8602, Japan}

\begin{document}

\maketitle

\begin{abstract}

We study false vacuum decay in a black hole (BH) spacetime with an angular momentum. 
Considering the false vacuum region described by a Kerr-de Sitter geometry, under the thin wall approximation, we can obtain the stationary configuration of the vacuum bubble seen from the outside false vacuum region without specifying the geometry inside the domain wall. 
Then, assuming the true vacuum region is described by a Kerr geometry, 
we can fix the mass and the spin parameter for the Kerr geometry by imposing the 1st junction conditions and conservation of the angular momentum. 
Although the assumption of the Kerr geometry inside the domain wall cannot be fully consistent with the 2nd junction conditions, we can roughly evaluate the error associated with this inconsistency by calculating the Brown-York 
quasi-local energy on the domain wall. 
Then the decay rate can be estimated by using the obtained parameters for the inside Kerr geometry and the Brown-York quasi-local energy. 
Our results support the statement that the BH spin suppresses the false vacuum decay in a BH spacetime.

\end{abstract}

\tableofcontents

\newpage 

\section{Introduction}
\label{Intro}

\begin{figure}[htbp]
   \begin{center}
 \includegraphics[clip,width=6cm]{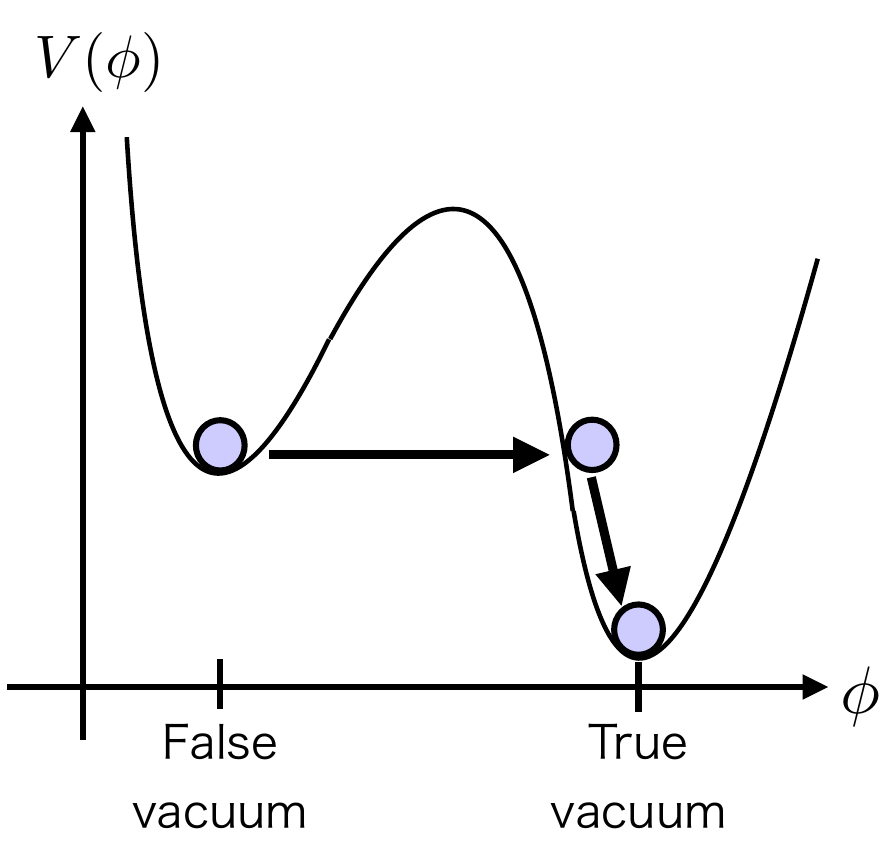}
    \caption{The schematic figure for the vacuum decay via quantum tunneling.} \label{fig:schem}
   \end{center}
 \end{figure}

\begin{figure}[htbp]
   \begin{center}
 \includegraphics[clip,width=14cm]{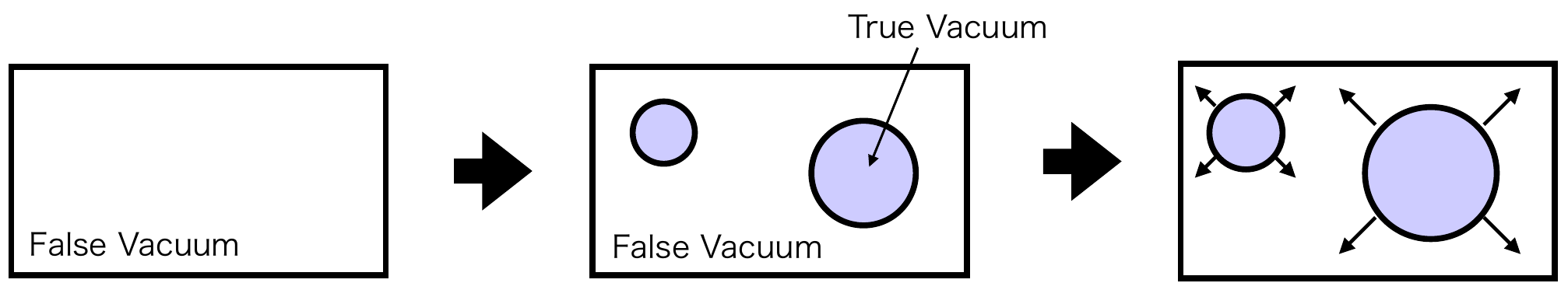}
    \caption{The schematic figure for vaccum phase transition.} \label{fig:spacetime}
   \end{center}
 \end{figure}

False vacuum decay is transition of a field from a metastable state (false vacuum) to a stable state (true vacuum) (Fig.~\ref{fig:schem}). 
The phenomenon is triggered by e.g. a thermal excitation and a quantum tunneling, which we focus on in this paper. 
Let us consider the transition in the spacetime filled up with a false vacuum field. 
The false vacuum decay occurs stochastically, and there will be some finite number of true vacuum regions, called vacuum bubbles. The bubbles expand typically and the true vacuum region will fill the spacetime eventually (vacuum phase transition) (Fig.~\ref{fig:spacetime}). 

False vacuum decay has been studied for more than 40 years since the pioneering work about vacuum decay in the flat background spacetime 
by Coleman~\cite{Coleman:1977py}. 
Coleman and De Luccia investigated vacuum decay in the maximal symmetric spacetime including the gravitational effect in~\cite{Coleman:1980aw}, and vacuum decay in a black hole (BH) spacetime was discussed in~\cite{Hiscock:1987hn} for the first time. 
In~\cite{Hiscock:1987hn}, it was reported that, for the Schwarzschild solution, namely, the static and spherically symmetric vacuum solution of the Einstein equations, a BH acts as a nucleation site for the decay. 
More recently, the transition in the Schwarzschild spacetime has been  
investigated in more detail~\cite{Gregory:2013hja,Burda:2015yfa}. 
According to the results in Refs.~\cite{Gregory:2013hja,Burda:2015yfa}, there are some cases where the existence of BHs raises the decay rate. 

Although there is no perfect understanding of a physical reason for the promoting effect of BHs, 
one of the most plausible explanations is that the effect is associated with thermal assistance~\cite{Mukaida:2017bgd}. 
In~\cite{Mukaida:2017bgd}, the authors give an interpretation of the effect in some limiting cases as follows: BHs are thermal sources of which temperatures are those Hawking temperatures, 
and tunneling with them is thermally assisted tunneling with the temperature. 
That is to say, the authors stated that the thermal radiation from the BHs enhance the vacuum decay rate. 
The thermal nature of BHs is regarded as a quantum gravity effect, so if we follow this interpretation, 
vacuum decay in BH spacetimes has quantum nature of gravity at a semi-classical level.  
In~\cite{Gregory:2013hja,Burda:2015yfa}, decreasing horizon area, which violates the area theorem in the classical gravity theory, was reported, so we can also regard this result as a consequence of quantum nature of gravity. 
These facts imply that vacuum decay with BHs may be a clue to quantum gravity theory.

False vacuum decay in BH spacetimes is getting attention for its applications in recent years. 
In 2012, the Higgs particle has been found~\cite{ATLAS:2012yve,CMS:2012qbp}. 
And then, it has been pointed out by several authors 
that the Higgs vacuum may be metastable and decay~\cite{Sher:1988mj,PhysRevD.40.613,Espinosa:1995se,Isidori:2001bm,Elias-Miro:2011sqh,Degrassi:2012ry,Buttazzo:2013uya,Kohri:2017iyl}. 
If the Higgs vacuum is metastable and there are many BHs to trigger the decay, the lifetime of the Higgs in our universe is shortened. 
Following such a perspective, in ~\cite{Dai:2019eei}, a constraint on the mass spectrum of primordial black holes (PBH) is provided. 

At the present time, there is much room to discuss vacuum decay with BHs. 
Besides poor understanding of physical meanings, we can point out that most of the related previous researches focused on spherically symmetric spacetimes. 
In general, realistic BHs are rotating, so discussions in spherically symmetric spacetimes are insufficient for applications. 
Therefore it would be meaningful to consider transitions in a rotating BH spacetime. 
Because of the lack of spherical symmetry, it is difficult to treat the problem as a one-dimensional problem as in Ref.~\cite{Gregory:2013hja}. 
Indeed, no robust way has been established to treat vacuum decay in a non-spherically symmetric BH spacetime. 
An analysis in the Kerr spacetime has been done in Ref.~\cite{Oshita:2019jan} for the first time with some assumptions such as no backreaction to the spacetime. 
According to Ref.~\cite{Oshita:2019jan}, increasing the BH spin decreases the decay rate. 
On the other hand, differently from the result in Ref.~\cite{Oshita:2019jan}, we found that the spin of the BH promotes the decay in the BTZ spacetime, which is a three-dimensional spacetime with an angular momentum~\cite{Saito:2021vut}.
The origin of this discrepancy has not been clarified yet. 
In this paper, we aim to obtain a deeper understanding by analyzing vacuum decay in the Kerr spacetime with different settings and assumptions from~\cite{Oshita:2019jan}. 
We take the thin wall approximation as many previous researches. 
In addition, we focus on the transition with a stationary bubble wall and evaluate the decay rate. 

This paper is organized as follows. 
In Sec.~\ref{Sum}, we briefly summarize the working hypotheses we make in this paper. 
In Sec.~\ref{config}, we show how to calculate the configuration of the bubble wall without knowing the geometry inside the bubble wall. 
In Sec.~\ref{1st}, assuming the inside geometry is given by a Kerr geometry, we fix the mass and spin parameters by demanding the continuity of the metric and the conservation of the angular momentum. 
In Sec.~\ref{2nd}, we discuss how we deal with the inconsistency in the 2nd junction conditions associated with the use of the Kerr geometry inside the bubble wall.  
We show the results in Sec.~\ref{Res} and discuss their implications in Sec.~\ref{Dis}.

Throughout this paper, we use the geometrized units in which both the speed of light and the Newton's gravitational constant are unity, $G=c=1$.

\section{Summary of the procedure to evaluate the decay rate}
\label{Sum}

As is stated in the introduction, no sophisticated treatment is known for the evaluation of the vacuum decay rate in a non-spherically symmetric spacetime. 
Our understanding is still far from a fully consistent treatment for the evaluation of the decay rate, and we need to introduce some working hypotheses. 
In order to avoid any confusion, let us first summarize our procedure to evaluate the decay rate in this section. 

According to Ref.~\cite{Coleman:1977py}, in a vacuum decay process via a quantum tunneling, the decay rate per unit spacetime volume $\Gamma$ is given by
\begin{align}
\Gamma\propto e^{-\mathcal{B}/\hbar}=e^{-(S_{E}-S_{E0})/\hbar},
\end{align}
where $S_{E}$ and $S_{E0}$ are the values of the Euclidean action 
for the Euclidean tunneling solution and the false vacuum solution, respectively, and $\mathcal{B}$ has been defined as
\begin{align}
\mathcal{B}:=S_{E}-S_{E0}. 
\end{align}
Apart from the overall factor, the decay rate can be mostly evaluated by the value of $\mathcal B$, and we focus on the evaluation of $\mathcal B$ in this paper.  

Throughout this paper, we rely on the thin wall approximation for the bubble wall between true and false vacuum regions, and do not consider any specific matter contents which realize the vacuum decay process. 
Although the thin wall approximation is one of our working hypotheses, 
at least in the case of the transition of a scalar field in the flat spacetime with a double well potential, the thin wall assumption would be justified under some conditions: a steeper potential peak between two minima and a smaller gap in the potential values in the local minima~\cite{Coleman:1977py}. 
This assumption is also adopted in many related papers, e.g., Refs.~\cite{Hiscock:1987hn,Gregory:2013hja,Mukaida:2017bgd}. 
Then we also regard the system we treat as composed of a false vacuum region, a true vacuum region, and a thin wall that mediates them satisfying the Israel junction conditions~\cite{Israel:1966rt}. 
In the neighborhood of the bubble wall, the system can be locally approximated by a planar domain wall system, and the intrinsic energy-momentum tensor $S_{ab}$ of the wall would be approximated by the pure tension-type: 
\begin{equation}
   S_{ab}=-\sigma h_{ab}, 
   \label{eq:pt}
\end{equation}
where $\sigma>0$ is the tension of the shell and $h_{ab}$ is the induced metric on the shell. 

For further simplification, we focus on the stationary configurations of the system, that is, we consider time-independent configurations. 
This assumption is not just a simplification. 
According to the results in Ref.\cite{Gregory:2013hja}, in the spherically symmetric case, the transition with the static shell gives the largest decay rate for a sufficiently large BH mass in the false vacuum region. 
Based on this fact, at least for slowly rotating cases, we may expect that the transition with the stationary shell also gives the maximum probability, and focus on such cases.  
Then, as is reviewed in Appendix~\ref{Action}, the difference of the Euclidean action $\mathcal{B}$ can be evaluated by computing the difference of the black hole horizon area before and after the transition.  
Since our purpose is reduced to the computation of the horizon areas in the stationary system, we perform all computations in the main text with the Lorentzian metric to avoid possible confusion. 

Our procedure to 
calculate the decay rate can be divided into the following 3 steps:
\begin{enumerate}
   \item For a given value of the tension $\sigma$, solving the shell equation of motion, we determine the bubble wall configuration seen
   from the outside Kerr-de Sitter region which is characterized by the mass parameter $M_+$, the spin parameter $a_+$, and the de Sitter length $l$. 
   This process can be done without specifying the geometry inside the wall once we assume the vanishing vacuum energy for the inside geometry as will be explained in Sec.~\ref{config}. 
   \item Assuming that the geometry inside the shell is given by the Kerr metric, we fix the mass parameter $M_-$ and the spin parameter $a_-$ by imposing the conservation of the angular momentum and the 1st junction conditions on the shell. 
   Then we can evaluate the decay rate by calculating the horizon area of the BH in the true vacuum region. 
   However, in this process, the assumption of the inside Kerr geometry cannot be justified without a non-spherical version of the Birkoff's theorem and leads to inconsistency in the 2nd junction conditions.
   \item We quantitatively evaluate the inconsistency in the 2nd junction conditions by calculating the Brown-York energy inside the shell with the Kerr geometry being the reference metric. 
   Since the evaluated Brown-York energy must be zero if the procedure is fully consistent, the value of the Brown-York energy allows us to evaluate an uncertainty for the decay rate. 
   Then we also show the decay rate with the correction evaluated from the Brown-York energy. 
\end{enumerate}
A more concrete description will be given in the following sections.

\section{Bubble configuration in a Kerr-de Sitter spacetime}
\label{config}

The Israel junction conditions consist of the 1st conditions
\begin{align}
 [h_{ab}]_{\pm}&=0, \label{eq:junc1} 
 \end{align}
and the 2nd conditions
\begin{align}
   [K_{ab}]_{\pm}&=-8\pi\left(S_{ab}-\frac{1}{2}h_{ab}S\right),
   \label{eq:junc2}
  \end{align}
where $K_{ab}$ is the extrinsic curvature on the shell. 
Here, for convenience, we labeled the quantities on the spacetime  
outside/inside of the bubble wall with the subscript $+/-$ and used the expression 
\begin{align}
   [A]_{\pm}&:=A_{+}-A_{-}.
\end{align}

Let us consider representing the EoM for the shell configuration
only with the geometrical quantities in the $+$-side. 
In order to do that, we use the junction conditions and the constraint equation on the shell. 
Let us consider the projection of the Einstein equations 
onto the normal direction to the shell (the Hamiltonian constraint),
\begin{align}
^3\mathcal{R}-K^2_{\pm}+K^{\pm}_{ab}K_{\pm}^{ab}=-16\pi T_{ab}n^{a}n^{b}. 
\label{eq:Hamilton}
\end{align}
Here, $T_{ab}$ is the energy-momentum of the $+$/$-$-side region,
and $n^{a}$ is the outward unit normal vector to the shell. 
By subtracting the $-$-side equation from the $+$-side equation, 
we obtain
\begin{align}
[K_{ab}]_{\pm}\bar{K}^{ab}-[K]_{\pm}\bar{K}=-8\pi[T_{ab}n^{a}n^{b}]_{\pm},
\end{align}
where we defined the expression
\begin{align}
\bar{A}&:=\frac{1}{2}(A_{+}+A_{-}).
\end{align}

By using Eq.\eqref{eq:junc2} and its trace
\begin{align}
[K]_{\pm}=4\pi S, \label{eq:trace}
\end{align}
we can obtain
\begin{align}
 S_{ab}\bar{K}^{ab}&=[T_{ab}n^{a}n^{b}]_{\pm}.
 \end{align}
In this paper, we assume that the false vacuum has positive vacuum energy, $T_{ab+}=-\frac{3}{8\pi l^2}g_{ab+}$, and vanishing vacuum energy for the true vacuum, $T_{ab-}n^{a}n^{b}=0$. 
Based on these assumptions, we obtain 
\begin{align}
[T_{ab}n^{a}n^{b}]_{\pm}
&=-\frac{3}{8\pi l^2}.
\end{align}
By using Eq.\eqref{eq:pt}, we get
\begin{align}
 S_{ab}K^{ab}
 &=-\sigma K,
 \end{align}
and 
 \begin{align}
K_{+}+K_{-}=\frac{3}{4\pi\sigma l^2}.
\label{eq:Ktp}
 \end{align}

By contracting Eq.\eqref{eq:junc2} with $S_{ab}$, we obtain
\begin{align}
K_{+}-K_{-}=-12\pi\sigma,
\label{eq:Ktm}
 \end{align}
 and summing Eqs.\eqref{eq:Ktp} and \eqref{eq:Ktm}, we can get 
\begin{align}
K_{+}=-6\pi\sigma+\frac{3}{8\pi\sigma l^2}. 
\label{eq:Km}
 \end{align}
This is the EoM for the shell configuration. 
Note that in order to obtain the EoM, the pure tension assumption is necessary, besides the Hamiltonian constraint and the 2nd junction conditions.

In this paper, we consider transition from a Kerr-de Sitter spacetime characterized by the BH mass $M_{+}$, spin $a_{+}$ and the de Sitter length $l$.
The metric in the Boyer-Lindquist coordinates is given by 
\begin{align}
   ds^2&=-\frac{1}{\rho^2}\qty(\Delta_{l}-a^2\sin^2\theta\Delta_{\theta})dt^2-\frac{2a\sin^2\theta}{\rho^2\Sigma_{l}}\qty[\Delta_{\theta}(r^2+a^2)-\Delta_{l}]dtd\phi+\frac{\rho^2}{\Delta_{l}}dr^2+\frac{\rho^2}{\Delta_{\theta}}d\theta^2 \nonumber \\
   &+\frac{\sin^2\theta}{\rho^2\Sigma_{l}}\qty[\Delta_{\theta}(r^2+a^2)^2-\Delta_{l}a^2\sin^2\theta]d\phi^2, \label{eq:KdS} \\
   \Delta_{l}&:=(r^2+a^2)\qty(1-\frac{r^2}{l^2})-2Mr, \label{eq:Dell} \\
   \Delta_{\theta}&:=1+\frac{a^2}{3l^2}\cos^2\theta, \label{eq:Delt} \\
   \Sigma_{l}&:=1+\frac{a^2}{3l^2}. \label{eq:Sigl}
\end{align}

Because of the axisymmetry, we can assume that the stationary shell is given as a hypersurface 
\begin{equation}
   \mathcal{W}:=\{(t_{\pm},r_{\pm},\theta_{\pm},\phi_{\pm})|r_{\pm}-R(\theta_{\pm})=0\},
   \label{eq:WK}
\end{equation}
and the unit normal vector to this surface is computed as
\begin{equation}
  n^{a}=\frac{1}{\sqrt{g^{rr}+g^{\theta\theta}\left(\partial_{\theta}R\right)^2}}\qty[g^{rr}(\partial_{r})^{a}-g^{\theta\theta}\partial_{\theta}R(\partial_{\theta})^{a}].
  \label{eq:na}
\end{equation}
By caluculating
\begin{align}
 K_{+}&=\nabla_{a}n^{a}|_{r_{+}=R(\theta_{+})}
 \label{eq:Kp}
\end{align}
from Eq.\eqref{eq:na} and putting it into Eq.\eqref{eq:Km}, we can obtain a second order differential equation 
for $R(\theta_+)$. 
By solving this equation, we can obtain the shell configuration seen by an observer outside
the shell.
We note that our treatment is fully consistent under the thin wall approximation thus far without specifying the inside geometry. 

\section{Determining the inside geometry}
\label{1st}

In the spherically symmetric case, the geometry can be restricted to the Schwarzschild-de Sitter family thanks to the Birkhoff's theorem. 
However, without spherical symmetry, we cannot easily specify the inside geometry by a known family of solutions. 
For a fully consistent treatment, we may have to solve the Euclidean Einstein field equations inside the shell with a boundary condition specified on the bubble wall. 
This is an extremely hard problem and, in this paper, we adopt an alternative way to avoid this difficulty. 
We simply assume that the inside geometry is given by the Kerr family described by \eqref{eq:KdS} with 
\begin{align}
   \Delta_{l}&\rightarrow r^2+a^2-2Mr, \\
   \Delta_{\theta}&\rightarrow1, \\
   \Sigma_{l}&\rightarrow1.
\end{align}
That is, the inside geometry is characterized by two parameters: the mass parameter $M_-$ and the spin parameter $a_-$. 
In our situation, because of the stationarity, the difference of the Euclidean action $\mathcal B$ can be calculated from the difference of the BH horizon areas. 
See Appendix~\ref{Action} for the derivation. 
BH horizon areas of 
a Kerr-de Sitter and Kerr spacetime 
are 
given by
\begin{align}
A_{\mathcal{H}_{+}}&=4\pi\frac{r^2_{+}+a^2_{+}}{1+\frac{a^2_{+}}{l^2}}, \\
A_{\mathcal{H}_{-}}&=8\pi M_{-}\left(M_{-}+2\sqrt{M^2_{-}-a^2_{-}}\right),
\label{eq:Ahm}
\end{align}
respectively, where $r_{+}$ is the outer event horizon radius for the Kerr-de Sitter. 

Once the values of $M_{+}$, $a_{+}$, and $l$ are fixed, we can compute $\mathcal B$ by knowing the values of the parameters $M_-$ and $a_-$. 
As is explained in the following subsections, we fix the values of $M_-$ and $a_-$ by using the angular momentum conservation and the 1st junction conditions. 
Since the forced inside geometry leads to inconsistency in the 2nd junction condition, we will evaluate the decay rate with the deviation associated with the inconsistency as will be explained in Sec.~\ref{2nd}. 
In this section, first, we explain how we fix the values of the parameters $M_-$ and $a_-$.  

 \subsection{Conserved angular momentum}

Since we are focusing on stationary and axisymmetric spacetimes, we may use a conserved charge associated with the symmetries to give a restriction on the possible parameter space of $M_-$ and $a_-$. 
More specifically, we can evaluate the quasi-local energy and angular momentum with the Komar integral~\cite{PhysRev.129.1873}.
However, as is explained in Appendix~\ref{Energy}, the energy conservation is not appropriate to be used in our setting.  
On the other hand, we can use the angular momentum conservation. 
Let us consider the Komar integral for the angular momentum given by 
\begin{align}
   J:=\frac{1}{16\pi}\oint_{r=R(\theta),\Sigma_{t}}\nabla^{a}\left(\frac{\partial}{\partial\phi}\right)^{b}dS_{ab}
   \end{align}
on the shell, where $dS_{ab}$ is given by  
\begin{align}
   dS_{ab}&=-2u_{[a}n_{b]}dS
\end{align}
with $u_a$ and $dS$ being the unit normal 1-form to the time slice $\Sigma_{t}$ and the 2-dim surface element of the shell
$\Sigma_{t}\cap\mathcal{W}$. 
Here, it should be noted that, since the spacetime is axisymmetric, the Killing vector $(\partial_\phi)^a$ is globally and uniquely defined with the period $2\pi$ of the coordinate $\phi$. 

By explicit calculation at just outside of the shell
$r_{+}=R(\theta_{+})$, 
we can compute the value of the angular momentum which is given by the following expression:
\begin{align}
J_{+}=-\int_{0}^{\pi}d\theta_{+} \frac{M_{+}a_{+}l^2 \sin ^3\theta_{+} \left(a_{+}^4-6R_{+}^4 -3a_{+}^2R_{+}^2-2
   a_{+}^2R_{+} (\partial_{\theta_{+}}R_{+}) \sin 2 \theta_{+} +a_{+}^2 \left(a_{+}^2-R_{+}^2\right) \cos 2 \theta \right)}{ \left(a_{+}^2+l^2\right)^2 \left(a_{+}^2+2R_{+}^2 +a_{+}^2 \cos 2\theta_{+}\right)^2}.
   \label{eq:Jp}
\end{align}
Here, we introduced the expression $R_{+}=R_{+}(\theta_{+})$ in order to stress that we are treating with the $+$-side quantities. 
Because of the pure tension assumption, the shell cannot have an intrinsic momentum and cannot carry the angular momentum. 
Therefore the conservation of the angular momentum leads to 
\begin{align}
M_{-}a_{-}=J_{+}. \label{eq:J}
\end{align}

\subsection{The 1st junction conditions}

The 1st junction conditions require the continuity of the induced metric on the shell. 
The line element on the shell can be written as
\begin{align}
ds^2_{\,\mathcal{W}}=h_{ab}^\pm dx_\pm^a dx_\pm^b=\qty[g_{tt}dt^2+2g_{t\phi}dtd\phi+g_{\phi\phi}d\phi^2+\qty{g_{\theta\theta}+g_{rr}\qty(\frac{dR}{d\theta})^2}d\theta^2]_{r_
{\pm}=R_{\pm}}
\end{align}
in the Boyer-Lindquist coordinates. 
Since the system is stationary and axisymmetric, the metric components do not depend on $t$ nor $\phi$, and a linear combination of the Killing vectors $(\partial_{t})^a$ and $(\partial_{\phi})^a$ is also a Killing vector. 
Therefore the first three terms describe a cylinder metric, and the continuity of the 2-dimensional part of the metric can be guaranteed by just imposing the equality of the circumferential radius. That is, regarding $g_{\phi\phi}$ as a function of $\theta$ and the radial coordinate $R(\theta)$ on the shell, we obtain 
\begin{equation}
   [g_{\phi\phi}(R(\theta),\theta)]_\pm=0\Leftrightarrow g^-_{\phi\phi}(R_-(\theta_-),\theta_-)-g^+_{\phi\phi}(R_+(\theta_+),\theta_+)=0.   
   \label{eq:eqph}
\end{equation}
Here, we introduced the notation $g^\pm_{\phi\phi}(R_\pm(\theta_\pm),\theta_\pm)$ for the metric component on the shell to emphasize that $g_{\phi\phi}$ depends only on $R=R(\theta)$ and $\theta$ there.
Since we already know $R_+(\theta_+)$, this gives a condition for $R_-(\theta_-)$. 

Because the unit vectors in the $+$ and $-$-side of the shell must be identical to each other on the shell, we obtain  
\begin{equation}
   \frac{d \theta_-}{d\theta_+}=\sqrt{\frac{h^+_{\theta\theta}}{h^-_{\theta\theta}}}. 
   \label{eq:eqth}
\end{equation}

Combining Eqs.\eqref{eq:eqph} and \eqref{eq:eqth}, we can obtain the following set of ordinary differential equations:
\begin{align}
   &\dv{\theta_{-}}{\theta_{+}}=\theta'_{-}(R_{-},\theta_{-},\theta_{+};M_{-}), 
   \label{eq:dthm}\\
   &\dv{R_{-}}{\theta_{+}}=R'_{-}(R_{-},\theta_{-},\theta'_{-},\theta_{+};M_{-}),
   \label{eq:dRm}
\end{align}
   where the prime denotes the derivative with respect to $\theta_{+}$. 
   We consider $M_+$, $a_+$ and $l$ as fixed parameters, and treat $R_+$ as a given function of $\theta_+$. 
   Note that the value of $a_-$ is given as a function of $M_-$ through \eqref{eq:Jp} and \eqref{eq:J}. 
We numerically solve them in the following way. 
First, we put a trial value of $M_{-}$ and set the initial value of  $\theta_{-}$ as $\theta_{-}(\theta_{+}=0)=0$. 
Then, we obtain the initial value of $R_{-}$ by demanding the 2nd junction condition \eqref{eq:junc2}:
\begin{align}
\left.\qty[K_{ab}u^au^b]_{\pm}\right|_{\theta_+=0}&=4\pi\sigma \label{eq:2ndtata} 
\end{align}
at $\theta=0$. 
Under this set up, we can solve and obtain $R_{-}=R_{-}(\theta_{-})$ depending on the value of $M_{-}$. 
Then we numerically find the value of $M_-$ which satisfies the regularity condition 
\begin{equation}
   R'_{-}|_{\theta_{-}=\frac{\pi}{2}}=0.    
\end{equation}

\section{Estimation of the deviation in the 2nd junction conditions}
\label{2nd}

As we discussed in the previous section, we can obtain $M_{-}$ and $a_{-}$ and compute $\mathcal{B}$ by solving the 1st conditions. 
However, we cannot fully impose the 2nd junction conditions 
\begin{align}
   K_{ab}^+-K_{ab}^-+4\pi\sigma h_{ab}=0.\label{eq:2ndtet}
\end{align}
The 2nd junction conditions are partially included in the conditions \eqref{eq:trace} and \eqref{eq:2ndtata}, and these are sufficient in the spherically symmetric case with the static configuration. 
However, without spherical symmetry, the other components are nontrivial and we inevitably suffer from the inconsistency of them with our procedure. 
Letting $K^0_{ab}$ be the extrinsic curvature for the inside Kerr geometry, we find
\begin{equation}
   \delta K_{ab}:=K_{ab}^--K_{ab}^0=K_{ab}^++4\pi\sigma h_{ab}-K_{ab}^0\neq 0. 
\end{equation}

The origin of the inconsistency can be regarded as the assumption of the inside Kerr geometry. 
In order to give a rough quantitative estimation for the correction of 
the decay rate due to the inconsistency, let us evaluate the Brown-York quasi-local energy \cite{Brown:1992br} of the region bounded by the shell with the inside Kerr geometry being the reference geometry. 
The Brown-York energy is defined by
\begin{align}
   E_{BY}&=-\frac{1}{8\pi}\int_{r=R(\theta),\Sigma_{t}}\qty(k-k_{0})dS, 
   \end{align}
   where $k$ is the extrinsic curvature on $\Sigma_{t}\cap\mathcal{W}$, and $k_{0}$ is the 
   fiducial extrinsic curvature to evaluate the energy. 
Apart from the use of the reference geometry, the Brown-York quasi-local energy can be evaluated only by the geometrical quantities on the boundary.
For our purpose, we may consider the inside Kerr geometry as the reference for $k_0$. That is, 
\begin{equation}
   k-k_0=\delta K+\delta K_{ab}u^{a}u^{b}.   
\end{equation}
Then, we may estimate the deviation $\delta M$ of the proper value $\bar M_-$ from $M_-$ evaluated on the boundary shell as 
\begin{equation}
   \delta M:=\bar M_- -M_-=E_{BY}. 
\end{equation}
We note again that, for $a_{+}$=0, the rest of the 2nd junction conditions \eqref{eq:2ndtet} are satisfied automatically and we obtain $\delta M=0$.
We can also expect $\delta M\ll M_-$ for $a_+\ll M_+$.

Since the corrected mass $\bar M_-$ can be given by 
   $\bar M_-=M_-+\delta M$, 
the corrected spin can be also evaluated as 
\begin{align}
   \bar a_{-}=\frac{J_{+}}{\bar M_-}=\frac{J_{+}}{M_{-}+\delta M},
 \end{align} 
due to the angular momentum conservation~\eqref{eq:J}. 
Although the proper geometry inside the shell would be different from the Kerr geometry and the horizon area would not be given by Eq.~\eqref{eq:Ahm}, we adopt the expression Eq.~\eqref{eq:Ahm} for the estimation of the deviation. 
Since we have two estimated values for $\mathcal B$ associated with $M_-$ and $\bar M_-$, hereafter, we express $\mathcal B$ with the argument as $\mathcal B(M_-)$ or $\mathcal B(\bar M_-)$.

A more speculative argument about this estimated value $\mathcal B(\bar M_-)$ is the following. 
Once the true vacuum region comes out in the Lorentzian spacetime and the shell expands far away from the BH, the BH may approach the Kerr BH. 
The mass for the Kerr BH may be expected to be bounded by the energy initially contained inside the shell, and the horizon area of the resultant Kerr BH is at most that with the mass $\bar M_-$. 
Namely, together with the area law, the estimated horizon area with the mass $\bar M_-$ may give an upper bound for the horizon area at the moment of the nucleation. 
Consequently, the estimated value of $\mathcal{B}(\bar M_-)$ would give a lower bound for $\mathcal B$. 

\section{Results}
\label{Res}

\subsection{Shell configuration}
\label{SC}

Before showing the results, we note that there should be a lower bound on the value of $M_{+}$ to have a stationary configuration.  
According to Ref.~\cite{Gregory:2013hja}, in the spherically symmetric case, if $M_{+}$ is smaller than the threshold $M_{C}$, there is no static transition. 
Here, $M_{C}$ is explicitly written as
\begin{align}
  \frac{M_{C}}{l}=\frac{1024\pi^3(\sigma l)^3}{27[16\pi^2(\sigma l)^2+1]^2}
\end{align}
for the spherically symmetric case.  
Conversely, if $M_{+}$ satisfies $M_{C}<M_{+}<M_{N}$, static transition exists, where $M_{N}/l=1/\sqrt{27}$ is the Nariai mass. 
Although we have no clear criteria for the case of rotating spacetimes, since the situation reduces to the static shell transition in the limit of $a_{+}\rightarrow0$, we restrict the parameter region to $M_{C}<M_{+}<M_{N}$. 

\begin{figure}[htbp]
    \begin{tabular}{cc}
      \centering
      \begin{minipage}[t]{0.45\hsize}
  \includegraphics[clip,width=7cm]{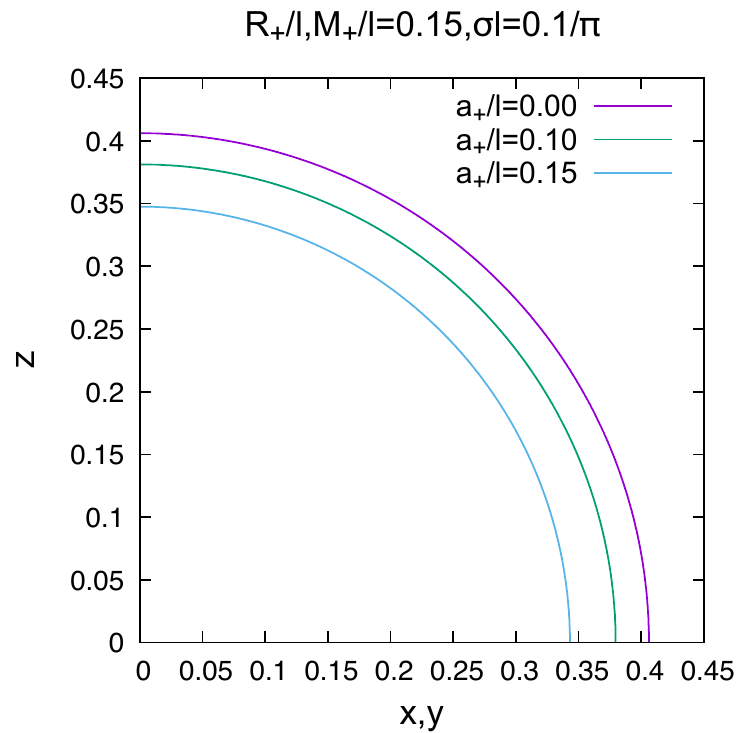}
      \end{minipage} &
      \begin{minipage}[t]{0.45\hsize}
      \centering
  \includegraphics[clip,width=7cm]{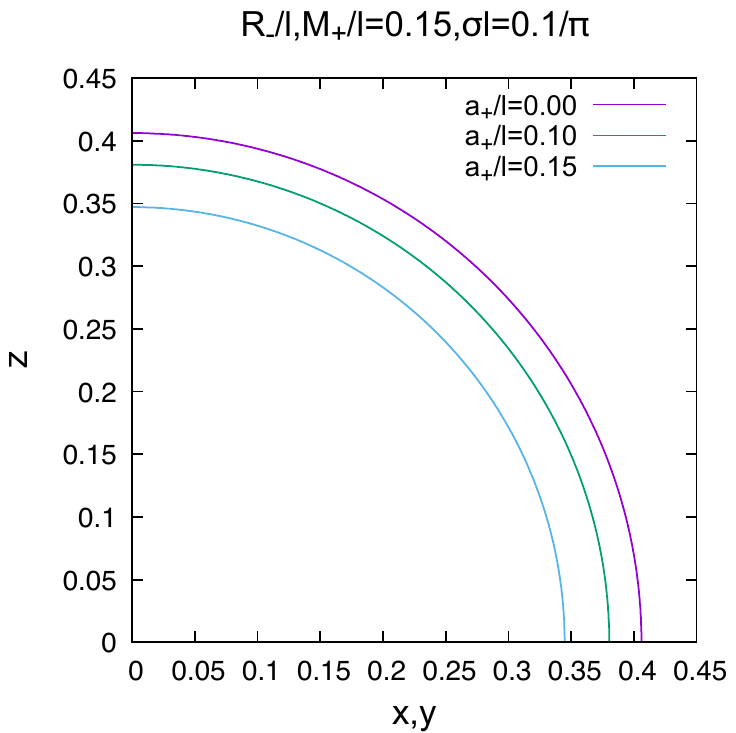}
      \end{minipage}
    \end{tabular}
     \caption{Configurations of the shell from the outside $R_+(\theta_+)$ (Left) and from the inside $R_-(\theta_-)$ (Right).} \label{fig:conf}
  \end{figure}

\begin{figure}[htbp]
   \begin{center}
 \includegraphics[clip,width=8cm]{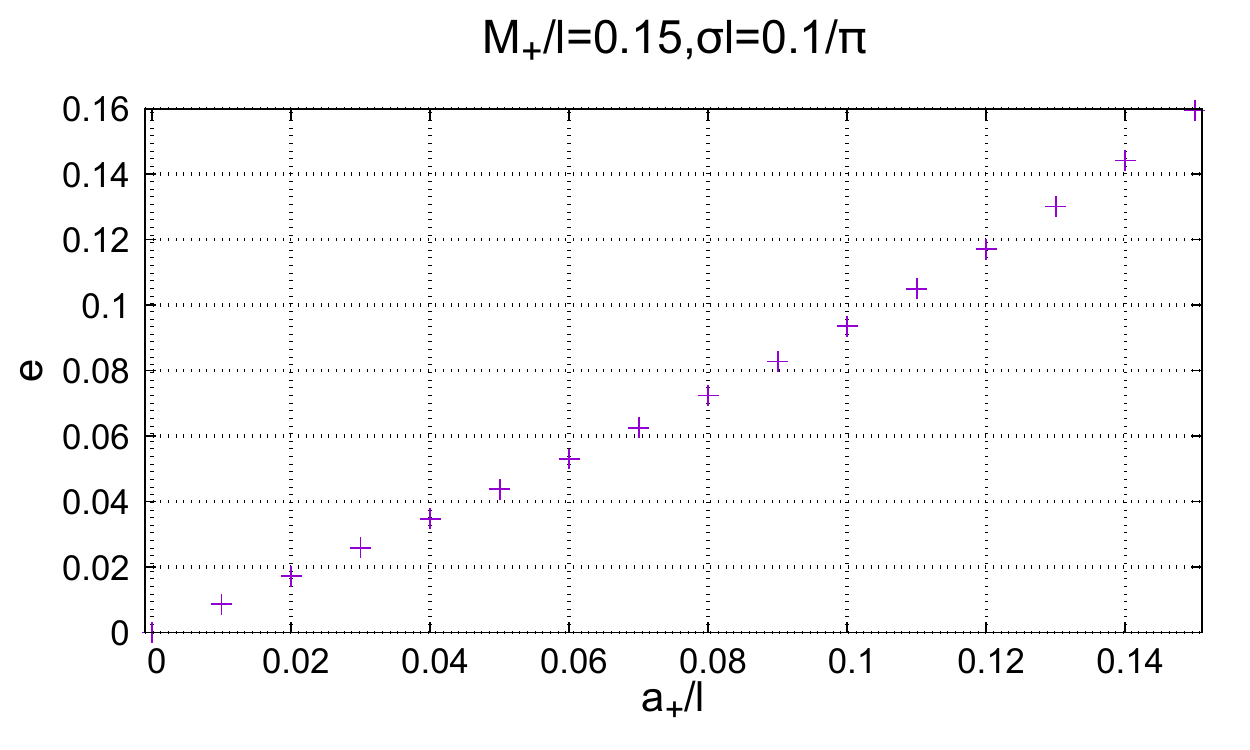}
    \caption{$a_{+}$ dependence of the eccentricity of $R_{+}(\theta_{+})$.} \label{fig:ellipse}
   \end{center}
 \end{figure}

Solving Eqs.~\eqref{eq:Kp}, \eqref{eq:dthm} and \eqref{eq:dRm}, we obtain the shell configurations $R_+(\theta_+)$ and $R_-(\theta_-)$ seen from the outside and inside observer, respectively, as depicted in Fig.~\ref{fig:conf}. 
First, we can see that the coordinate radius decreases with $a_{+}$ for both cases.  
In addition, the shell configuration gets more distorted for a larger value of $a_+$ as is explicitly shown in Fig.~\ref{fig:ellipse}, where the eccentricity obtained by fitting $R_{+}(\theta_{+})$ with an ellipsoid is depicted as a function of $a_+$. 
We can also confirm that the shell is spherical for $a_{+}=0$ as is expected. 

 \subsection{The Mass and spin of the BH after the transition}
\label{Ma}

\begin{figure}[htbp]
   \begin{minipage}[t]{0.45\hsize}
    \centering
    \includegraphics[clip,width=8cm]{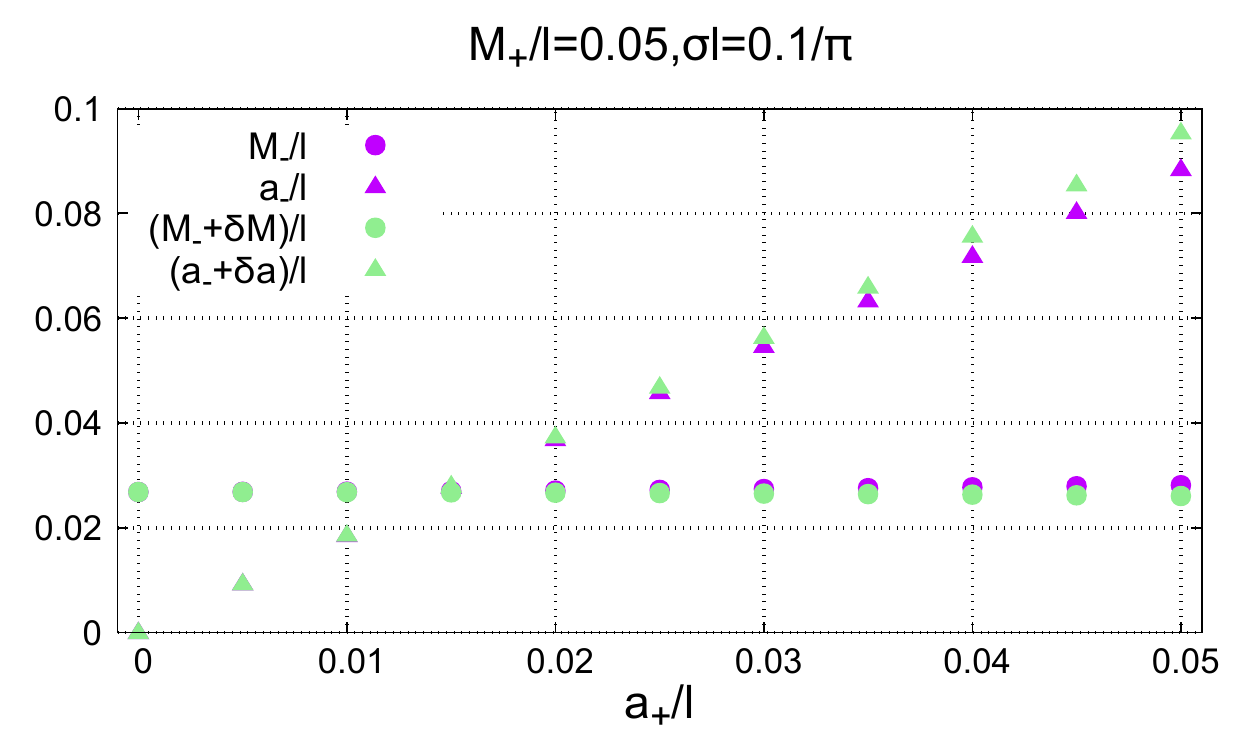}
    \subcaption*{}\label{fig:0.05Ma}
   \end{minipage}
   \begin{minipage}[t]{0.45\hsize}
    \centering
    \includegraphics[clip,width=8cm]{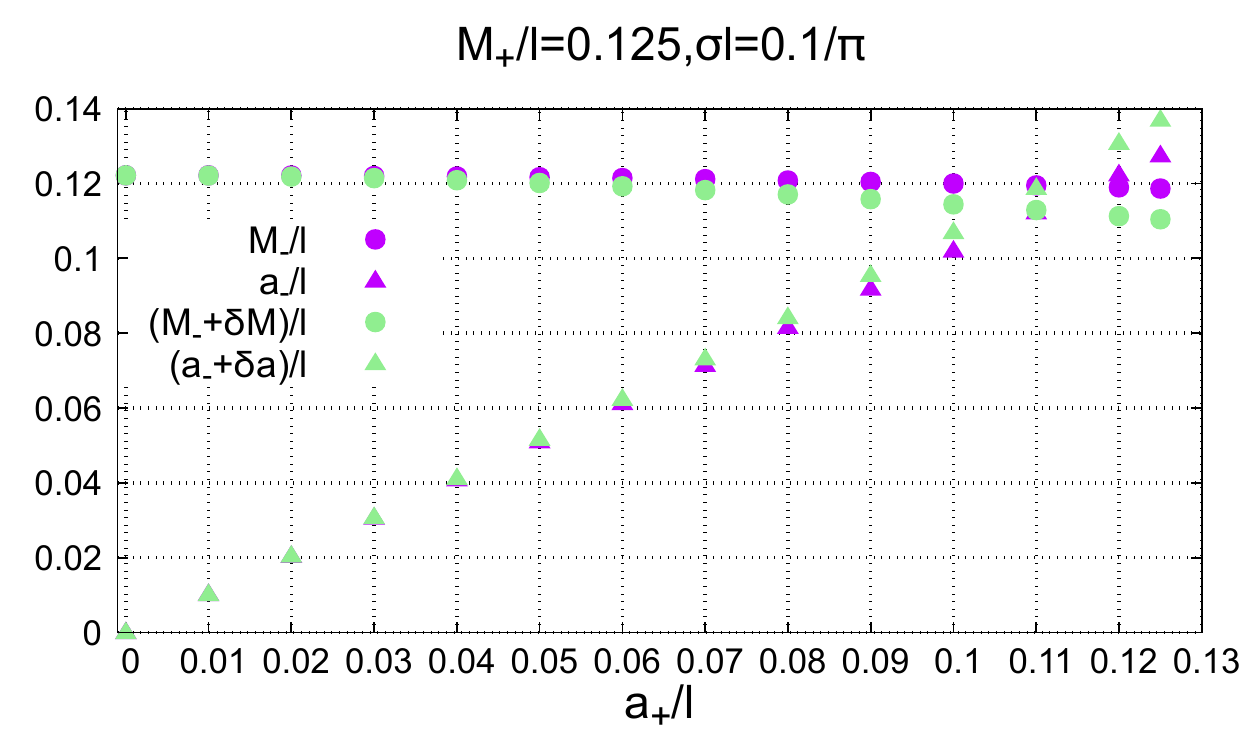}
    \subcaption*{}\label{fig:0.125Ma}
   \end{minipage}\\
   \caption{The $a_{+}$ dependence of $M_{-}$ and $a_{-}$ }\label{fig:Mam}
  \end{figure}

\begin{figure}[htbp]
   \begin{minipage}[t]{0.45\hsize}
    \centering
    \includegraphics[clip,width=8cm]{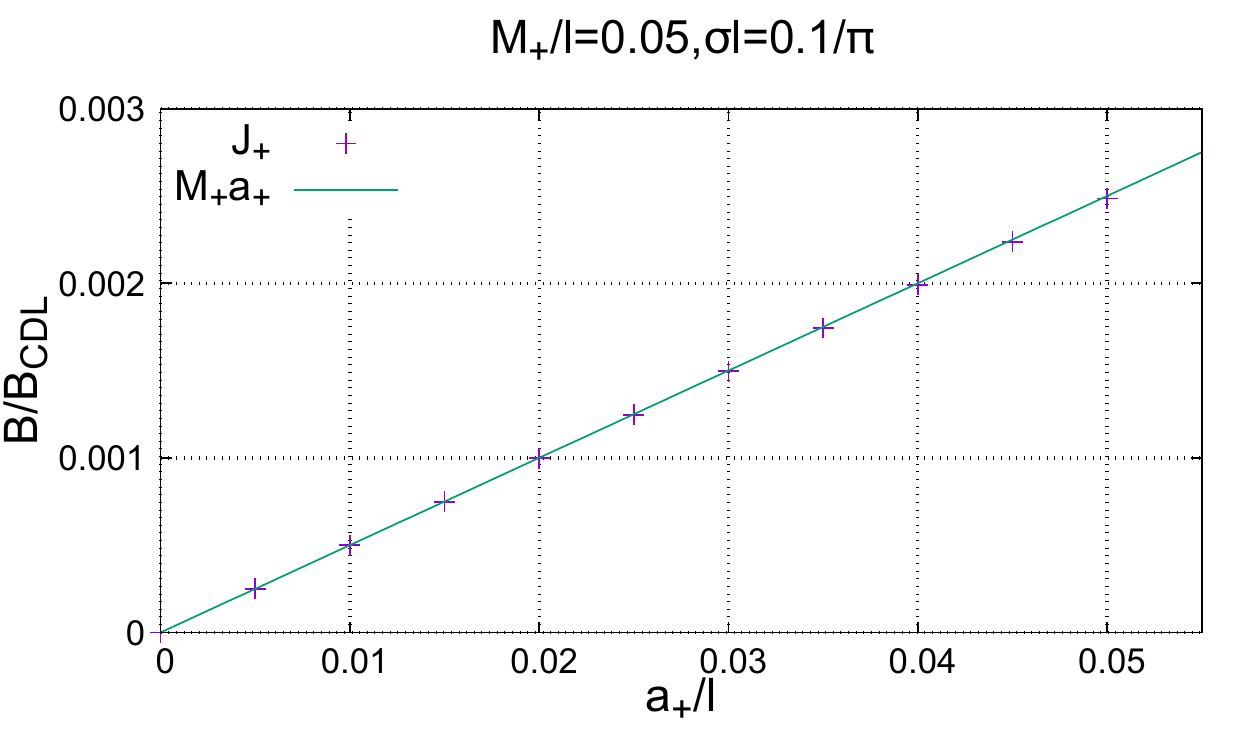}
    \subcaption*{}\label{fig: 0.05J}
   \end{minipage}
   \begin{minipage}[t]{0.45\hsize}
    \centering
    \includegraphics[clip,width=8cm]{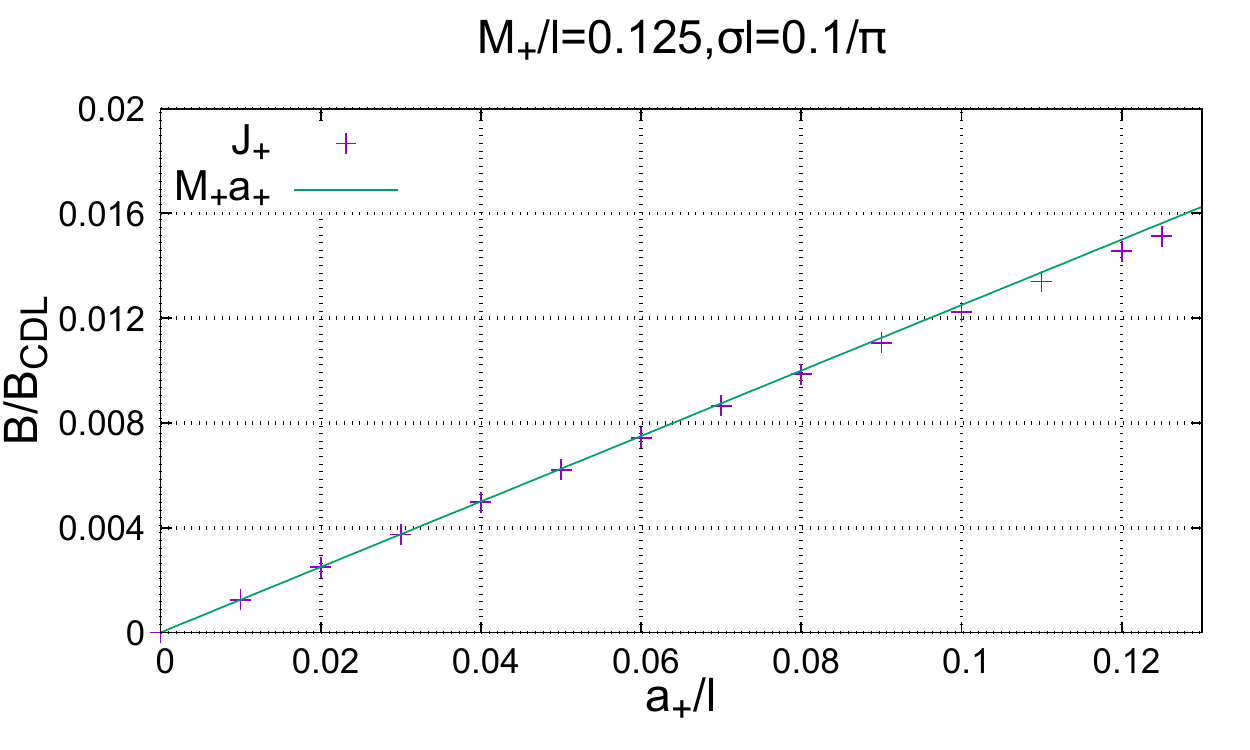}
    \subcaption*{} \label{fig:A0.125J}
   \end{minipage}
   \caption{The value of $J_{+}$ and $M_{+}a_{+}$ }\label{fig:J}
  \end{figure}

Through the procedure stated in the previous sections, we obtain the values of $M_-$, $a_-$ and $\delta M$. 
We plot those values as functions of $a_+/l$ for $M_+/l=0.05$ and $0.125$ with $\sigma l=0.1/\pi$ in Fig.~\ref{fig:Mam}. 
From Fig.~\ref{fig:Mam}, we can see that the values of $\delta M$ are always negative, consequently we have $\bar M_-<M_-$ and $\bar a_->a_-$. 
While $\delta M=0$ for $a_+=0$, we have nonzero $\delta M$ for $a_+>0$ and its absolute value increases with $a_+$.
This fact implies that the assumption of the Kerr geometry inside the shell more significantly conflicts with the 2nd junction conditions for a larger value of $a_+$.  
In the both cases for $M_+/l=0.05$ and $0.125$, we find an over-spinning Kerr geometry, namely, $a_->M_-$ for a relatively larger value of $a_+/l$. 
Since the deviation $\delta M$ is relatively smaller than the $a_+$-dependence of $M_-$ and $a_-$, we also find $\bar a_- > \bar M_-$ for a relatively larger value of $a_+/l$. 
We note that, before the transition, the central BH is a Kerr-de Sitter BH with $a_+<M_+$, and the singularity is hidden inside the horizon. 
Nevertheless, as a consequence of the matching conditions on the shell, we obtain the over-spinning BH. 
This consequence can be easily understood by considering the angular momentum conservation as follows. 
First, we can find that the resultant value of the quasi-local angular momentum $J_{+}$ can be approximately given by $M_{+}a_{+}$ (see Fig.~\ref{fig:J}). 
Then, by using the angular momentum conservation, we obtain
 \begin{align}
   &M_{-}a_{-}=J_{+}\simeq M_{+}a_{+},  \nonumber \\
   &\therefore \frac{a_{-}}{M_-}\simeq\frac{M_{+}a_{+}}{M_{-}^2}\leq\frac{M_{+}^2}{M_{-}^2}. 
\end{align}
Therefore we would find $\bar a_-\simeq a_->M_-\simeq \bar M_-$ for a sufficiently large $a_+$ if $M_+>M_-$ 
\footnote{According to Ref.~\cite{Gregory:2013hja}, for the static shell transitions in Schwarzschild spacetimes, 
the value of $M_-$ is smaller than $M_+$ unless the value of $M_+$ is large enough. 
For example, we obtain $M_{-}=M_{+}$ for $\sigma l =0.1/\pi$ and $M_{+}/l=0.1327$.}.
Since we cannot estimate the transition rate for the case $a_->M_-$ because of the existence of the naked singularity, we will not further discuss these cases in this paper.

\subsection{Decay rate}
\label{Dr}

\begin{figure}[htbp]
   \begin{center}
    \includegraphics[clip,width=8cm]{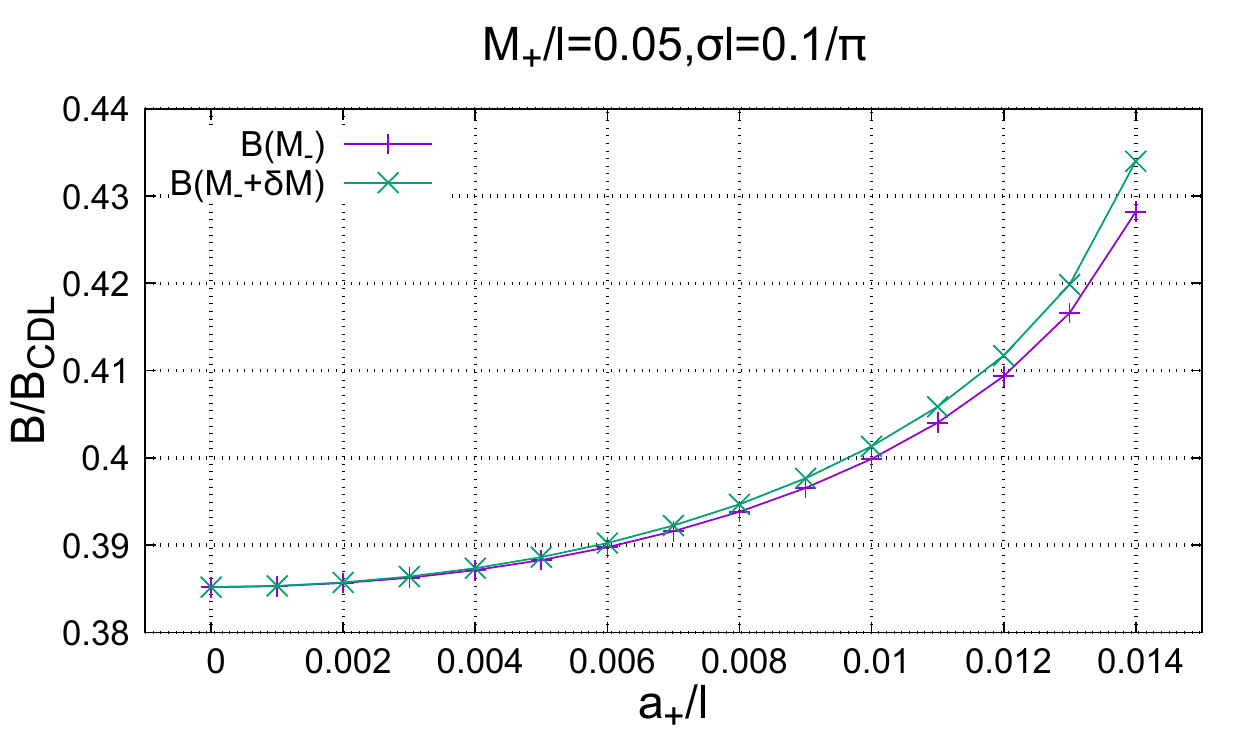}
  \includegraphics[clip,width=8cm]{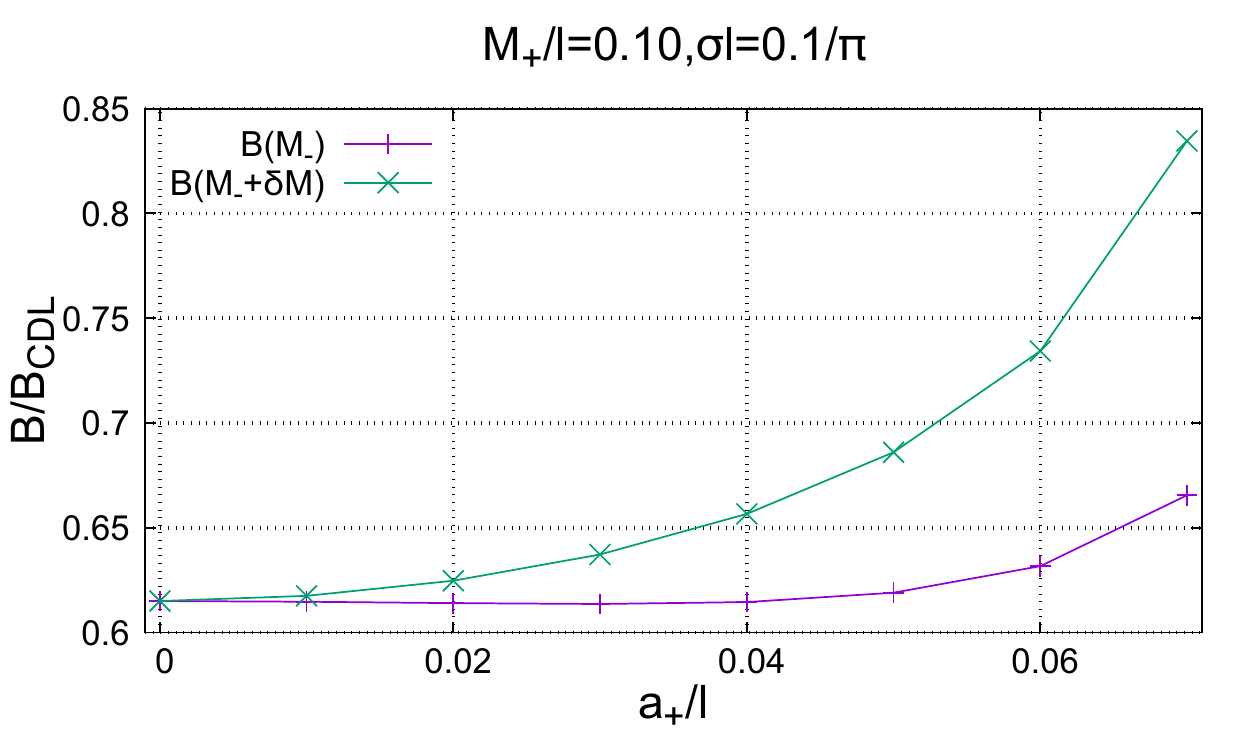}
    \includegraphics[clip,width=8cm]{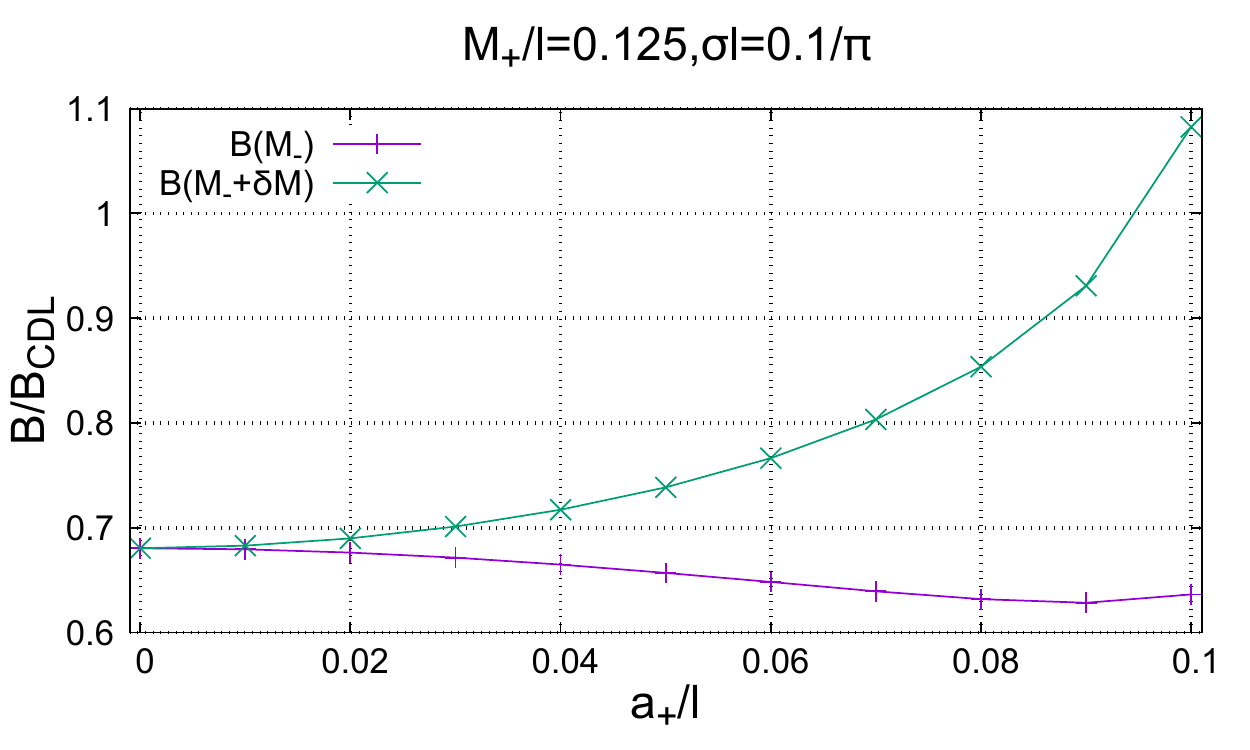}
   \caption{The $a_{+}$ dependence of $\mathcal{B}$}\label{fig:B}
 \end{center}
  \end{figure}

Finally, we show the $a_{+}$ dependence of the difference of the Euclidean action $\mathcal{B}$. 
We normalize $\mathcal{B}$ with $\mathcal{B}_{CDL}$, the value in the maximally symmetric spacetimes~\cite{Coleman:1980aw}. 
The results are depicted in Fig.\ref{fig:B} for $M_+/l=0.05$, $0.1$ and  $0.125$ with $\sigma l=0.1/\pi$. 
As is stated in Sec.~\ref{Ma}, since $\delta M<0$, we obtain $\mathcal B(\bar M_-)>\mathcal B(M_-)$ and $\mathcal B(\bar M_-)-\mathcal B(M_-)$ increases with $a_+/l$. 
In the case of $M_{+}/l=0.05$ and $\sigma l=0.1/\pi$, both of $\mathcal{B}(M_-)$ and $\mathcal B(\bar M_-)$ increase with $\tilde{a}_{+}$. 
By recalling that the decay rate is proportional to $\exp(-\mathcal{B}/\hbar)$, this result means that the spin decreases the rate. 
On the other hand, for the $M_{+}/l=0.10$ and $0.125$ cases, $\mathcal{B}(M_-)$ decrease with $a_{+}/l$ for $a_{+}/l \ll 1$, and start to increase from a certain value of $a/l$. 
However, as is explicitly shown in the figures, $\mathcal B(\bar M_-)$ is always an increasing function of $a_+$. 
Therefore, for the $M_{+}/l=0.10$, $0.125$ cases, we cannot give any concrete conclusion. 
Nevertheless, if we accept the speculative argument stated in the last part of Sec.~\ref{2nd}, $\mathcal B(\bar M_-)$ gives the lower bound for $\mathcal B$, which gives an upper bound for the decay rate.
Since the upper bound for the decay rate is smaller than the decay rate for $a_+=0$, we reach the same conclusion as for $M_{+}/l=0.05$, that is, the spin decreases the decay rate.

\section{Summary and discussion}
\label{Dis}

In this paper, we have discussed false vacuum decay in a Kerr-de Sitter spacetime. 
Because of the lack of spherical symmetry, a fully consistent analysis is very hard. Instead of trying to make a self-consistent procedure, we performed the tricky analysis briefly described below. 

First, we consider a Kerr-de Sitter geometry 
as the false vacuum region. 
Assuming that the bubble wall is described by the stationary thin shell with the equation of state being the pure tension-type, we can fix the configuration of the bubble wall seen from the outside region following the junction conditions without knowing the inside geometry. 
Under these assumptions, the decay rate can be calculated once we obtain horizon area of a BH in the true vacuum region. 
However, the inside geometry is not necessarily given by a Kerr geometry because of the lack of the Birkhoff's theorem which only works with spherical symmetry. 
Nevertheless, we can find a Kerr geometry that fits the shell imposing the angular momentum conservation and the 1st junction conditions. 
That is, we can obtain the mass parameter $M_-$ and the spin parameter $a_-$ for the inside geometry. 
Although we expect that the decay rate estimated from the fitted Kerr geometry $\Gamma(M_-)$ gives a good estimation at least for a relatively small Kerr parameter, we can go further by evaluating the deviation associated with the inconsistency in the 2nd junction conditions. 
Since we do not fully take into account the 2nd junction conditions when we look for the inside Kerr geometry, there should be inconsistency in the 2nd junction conditions. 
In order to evaluate the inconsistency, we calculate the Brown-York energy $\delta M$ with the inside Kerr geometry being the reference geometry. 
The value of $\delta M$ can be regarded as an error associated with the inconsistency in the 2nd junction conditions if we accept the inside Kerr geometry as an approximation for the unknown true geometry. 
Then we may estimate the corrected mass parameter inside geometry by $\bar{M}_{-}=M_-+\delta M$. 
In this sense, the decay rate is estimated at around the values $\Gamma(M_-)$ and $\Gamma(\bar{M}_{-})$ with an expected magnitude of the uncertainly $\sim |\Gamma(M_-)-\Gamma(\bar M_-)|$.  

There might be another interpretation about the mass parameter $\bar{M}_{-}$. 
Once the bubble is nucleated in the Lorentzian spacetime, and the bubble wall expands, the inside geometry would approach a Kerr spacetime. 
Then we naively expect that the mass inside the shell $\bar{M}_{-}$ would be carried by the resultant Kerr BH. 
Because of the area law in the classical gravity, the horizon area of the inside BH increases with time in the Lorentzian spacetime. 
Thus the horizon area estimated by the Kerr geometry with the mass parameter $\bar{M}_{-}$ should give an upper bound for the horizon area at the moment of the nucleation. 
Although this interpretation is not very solid and speculative because of the ambiguity of the quasi-local gravitational energy,  
once we accept it, the decay rate $\Gamma(\bar{M}_{-})$ gives an upper limit. 

The results are summarized as follows. 
For a relatively small mass parameter of a Kerr-de Sitter spacetime for the false vacuum region, letting $\Gamma_{\rm Sph}$ denote the decay rate for the spherically symmetric case, we obtain $\Gamma(\bar{M}_{-})<\Gamma(M_-)<\Gamma_{\rm Sph}$. 
That is, we can safely say that the decay rate is suppressed by the spin of the BH for a relatively small mass parameter. 
For a relatively large value of the mass parameter, we obtain $\Gamma(\bar{M}_{-})<\Gamma_{\rm Sph}<\Gamma(M_-)$. 
Therefore, adopting $\delta M$ as an error and the estimation of the decay rate as around $\Gamma(M_-)$ and $\Gamma(\bar{M}_{-})$, we cannot make any concrete conclusion. 
However, if we accept $\Gamma(\bar{M}_{-})$ as the upper bound for the decay rate, we may obtain the same conclusion as for the case of a relatively small mass parameter, that is, the spin of the BH suppresses the decay. 

Let us compare these results with the previous research in the Kerr spacetime~\cite{Oshita:2019jan}. 
In Ref.~\cite{Oshita:2019jan}, the authors focused on the small spin limit $a_{+}\ll l$ and assumed that there is no backreaction to the BH parameters. 
Moreover, they assumed that the shell has the anisotropic pressures $p_{\theta}$ and $p_{\psi}$, 
and treated the system in the form of the potential problem as often used in the spherically symmetric case.
As a result, they stated that the spin of the black hole decreases the decay rate as is consistent with our results.
On the other hand, according to our work in the BTZ spacetime~\cite{Saito:2021vut}, the spin in the three-dimensional rotating spacetime promotes the decay.
The reason for this difference has not yet been clarified.   
One possible cause is dimensional dependence, which would be investigated in future works.
In this paper, we performed a tricky procedure to avoid difficulties associated with lack of the spherical symmetry and took a rough estimation.
In order to get a concrete answer, we need to perform a totally consistent analysis which may require the construction of the inside geometry by solving a set of elliptic equations for geometrical quantities. 
Throughout the paper, we have used the thin wall approximation and assumed that the shell has pure tension. 
Relaxing these assumptions would be a valuable future work. 

\section*{Acknowledgements}

The author (DS) would like to take this opportunity to thank the “Nagoya University Interdisciplinary Frontier Fellowship” supported by Nagoya University and JST, the establishment of university fellowships towards the creation of science technology innovation, Grant Number JPMJFS2120.
This work was supported in part by JSPS KAKENHI Grant Nos.
JP19H01895 (CY), JP20H05850 (CY) and JP20H05853 (CY). 

\appendix

\section{Euclidean action and decay rate}
\label{Action}

In this subsection, by following Ref.~\cite{Gregory:2013hja}, we calculate $S_{E}$, 
the Euclidean action of the system, and derive the specific form in the case of a stationary configuration. 
Throughout this section, all computations will be done with the Euclidean signature. 

\begin{figure}[htbp]
   \begin{center}
 \includegraphics[clip,width=8cm]{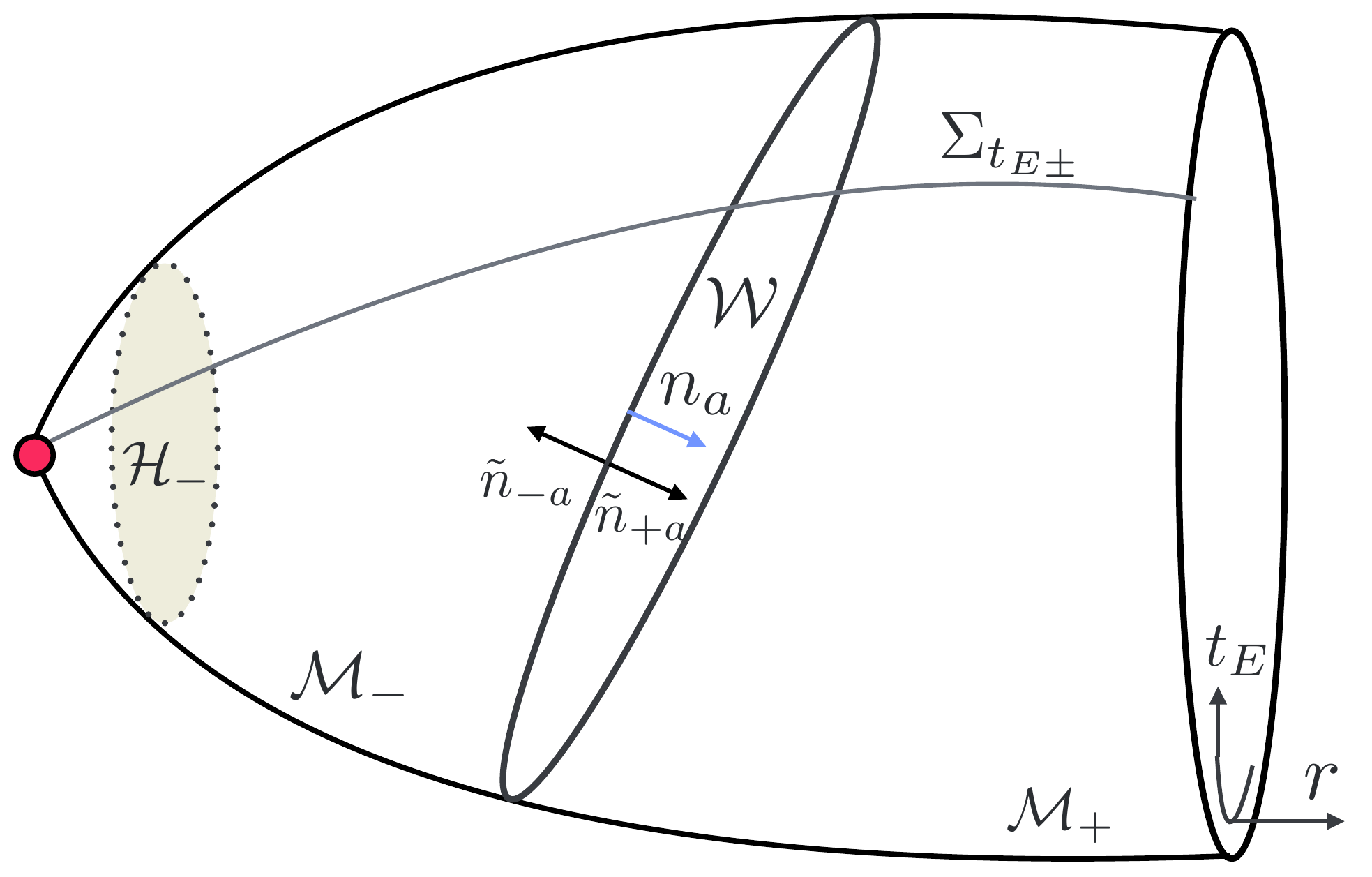}
    \caption{Schematic picture of Euclidean spacetime.} \label{fig:Eaction}
   \end{center}
 \end{figure}

We can divide $S_{E}$ into 
four parts as follows (Fig.~\ref{fig:Eaction}):
\begin{align}
S_{E}=S_{\mathcal{H}}+S_{\mathcal{M}_{-}}+S_{\mathcal{M}_{+}}+S_{\mathcal{W}},
\end{align}
where $S_{\mathcal{H}}$ is the contribution from conical defects.  
By taking 
appropriate coordinates near the 
defects~\cite{Gregory:2013hja},  
we can obtain 
\begin{align}
S_{\mathcal{H}}=-\frac{A_{\mathcal{H}_{-}}}{4}-\frac{A_{c}}{4}.
\end{align}
where $A_{\mathcal{H}_{-}}$ and $A_{c}$ are the BH event horizon area and cosmological horizon area, respectively.  

$S_{\mathcal{W}}$ is the contribution of the matter field on the shell $\mathcal{W}$. 
Under the thin wall approximation and the pure tension assumption, we obtain 
\begin{align}
S_{\mathcal{W}}&=-\int_{\mathcal{W}}dW\int_{R-0}^{R+0}dr{\mathcal{L}_{m}}^{(E)} \nonumber \\
&\simeq\int_{\mathcal{W}}dW\int_{R-0}^{R+0}dr\sigma\delta(r-R(\tau_E)) \nonumber \\
&=\int_{\mathcal{W}}dW\sigma, \label{eq:SW}
\end{align}
where $dW$ is the spacetime volume element and the integration is over the wall and Euclidean time.
 
Here, $r$ is the Gaussian normal coordinate to the wall. 

$S_{\mathcal{M}_{\pm}}$ is the contributions from the $+/-$-side of the domain wall with the boundary terms, which are written as
\begin{align}
S_{\mathcal{M}_{\pm}}&=-\int_{\mathcal{M}_{\pm}}dV\left(\frac{1}{16\pi}\mathcal{R}^{(E)}+\mathcal{L}_{m}^{(E)}\right)+\frac{1}{8\pi}\int_{\partial\mathcal{M}_{\pm}}dW\tilde{K}_{E\pm},
\end{align}
where $dV$ is the spacetime volume element. 
By taking the 3+1 decomposition
\begin{align}
   \mathcal{R}^{(E)}=^3\mathcal{R}^{(E)}-\tilde{K}^2_{E\pm}+\tilde{K}_{E\pm ab}\tilde{K}_{E\pm}^{ab}+2\nabla_{a}\qty(\tilde{u}_{\pm}^{b}\nabla_{b}\tilde{u}_{\pm}^{a})-2\nabla_{b}\qty(\tilde{u}_{\pm}^{a}\nabla_{a}\tilde{u}_{\pm}^{b}),
   \end{align}
on the constant (Euclidean) time slice $\Sigma_{t_{E\pm}}$, we can write them as 
\begin{align}
S_{\mathcal{M}_{\pm}}&=-\frac{1}{16\pi}\int dV\left(^3\mathcal{R}^{(E)}-\tilde{K}^2_{E\pm}+\tilde{K}_{E\pm ab}\tilde{K}_{E\pm}^{ab}+16\pi\mathcal{L}_{m}^{(E)}\right)  \nonumber \\
&+\frac{1}{8\pi}\int_{\mathcal{W}}dW\tilde{K}_{E\pm}+\frac{1}{8\pi}\int_{\mathcal{W}}dW\tilde{n}_{\pm a}\tilde{u}^{b}_{\pm}\nabla_{b}\tilde{u}^{a}_{\pm} \label{eq:ADM}.
\end{align}
Here, $\tilde{u}^{a}_{\pm}$ is the unit normal to $\Sigma_{t_{E\pm}}$, and $\tilde{n}_{\pm a}=\pm n_{a}$ is the inwarding unit normal vactor to $\mathcal{W}$. $\tilde{K}_{E\pm}$ is the Euclidean extrinsic curvature, which is related to $K_{\pm}$ via $K_{\pm}=\pm\tilde{K}_{E\pm}$.
The first line in Eq.\eqref{eq:ADM} vanishes due to the Hamiltonian constraint and the Killing symmetry~\cite{Gregory:2013hja}.
Then we obtain  
\begin{align}
S_{\mathcal{M}_{+}}+S_{\mathcal{M}_{-}}&=\frac{1}{8\pi}\int_{\mathcal{W}}dW(K_{+}-K_{-})+\frac{1}{8\pi}\int_{\mathcal{W}}dW(n_{a}\tilde{u}^{b}_{+}\nabla_{b}\tilde{u}^{a}_{+}-n_{a}\tilde{u}^{b}_{-}\nabla_{b}\tilde{u}^{a}_{-}) .
\label{eq:SMM} \end{align}
The second term can be rewritten as 
\begin{align}
\frac{1}{8\pi}\int_{\mathcal{W}}dW(n_{a}\tilde{u}^{b}_{+}\nabla_{b}\tilde{u}^{a}_{+}-n_{a}\tilde{u}^{b}_{-}\nabla_{b}\tilde{u}^{a}_{-})
&=-\frac{1}{8\pi}\int_{\mathcal{W}}dW(\tilde{u}^{a}_{+}\tilde{u}^{b}_{+}\nabla_{b}n_{a}-\tilde{u}^{a}_{-}\tilde{u}^{b}_{-}\nabla_{b}n_{a}) \nonumber \\
&=-\frac{1}{8\pi}\int_{\mathcal{W}}dW(K_{+uu}-K_{-uu}),
\end{align}
where $K_{\pm uu}:=\tilde{u}^{a}_{\pm}\tilde{u}^{b}_{\pm}\nabla_{b}n_{a}$ and in the first equality, we used the relation $\tilde{u}^{a}_{\pm}\tilde{n}_{a}=0$. 
By using the 2nd conditions
\begin{align}
K_{+}-K_{-}&=-12\pi\sigma, \\
K_{+uu}-K_{-uu}&=-4\pi\sigma h_{uu} =-4\pi\sigma,
 \end{align}
 we obtain 
\begin{align}
S_{\mathcal{M}_{+}}+S_{\mathcal{M}_{-}}&=\frac{1}{8\pi}\int_{\mathcal{W}}dW(-12\pi\sigma)+\frac{1}{8\pi}\int_{\mathcal{W}}dW(4\pi\sigma) \nonumber\\
&=\int_{\mathcal{W}}dW(-\sigma). \label{eq:SM}
\end{align}
Here, $h_{uu}=h_{ab}\tilde{u}^{a}_{\pm}\tilde{u}^{b}_{\pm}=(g_{ab}-n_{a}n_{b})\tilde{u}^{a}_{\pm}\tilde{u}^{b}_{\pm}=1$. 
Note that $\tilde{u}^{a}_{\pm}\tilde{u}_{\pm a}=1$ in the Euclidean spacetimes.
Using Eq.\eqref{eq:SW} and Eq.\eqref{eq:SM}, we obtain
\begin{align}
   S_{\mathcal{M}_{+}}+S_{\mathcal{M}_{-}}+S_{\mathcal{W}}&=\int_{\mathcal{W}}dW(\sigma-\sigma)=0,
   \end{align}
and we can see that only $S_{\mathcal{H}}$ contributes to $\mathcal{B}$.

Eventually, $\mathcal{B}$ reduces to
\begin{align}
\mathcal{B}=\frac{1}{4}(A_{\mathcal{H}_{+}}-A_{\mathcal{H}_{-}}) .
\end{align}
From the expression, we can see that the exponent of the decay rate depends only on the BH horizon areas. 
Note that the stationarity of the shell, $\tilde{u}^{a}_{\pm}n_{a}=0$ is necessary to obtain this result.

\section{No use of energy conservation}
\label{Energy}

Because of the stationarity of the system, one might think that the energy conservation law is useful with the use of the Komar integral~\cite{PhysRev.129.1873}
\begin{align}
   E_{\rm Kom}:=-\frac{1}{8\pi}\int_{r=R(\theta),\Sigma_{t}}\nabla^{a}\xi^b_{(t)}dS_{ab},  \label{eq:QLE}
   \end{align}
   where $\xi^a_{(t)}$ is a timelike Killing vector. 
However $E_{\rm Kom}$ is ill-defined for the following reason. 
Let us define the timelike Killing vector associated with the false vacuum as
\begin{align}
    \xi_{(t)+}^{a}=\left(\frac{\partial}{\partial t_+}\right)^a.
\end{align}
   For the quasi-local energy to be well-defined, the Killing vector must be defined globally in the system. 
   Therefore, because of the normalization, the Killing vector in the true vacuum must be 
\begin{align}
   \xi_{(t)-}^{a}=\sqrt{\frac{-g^{+}_{tt}(\theta)}{-g^{-}_{tt}(\theta)}}\left(\frac{\partial}{\partial t_-}\right)^a
   \end{align}
   on the bubble wall. 
However, the normalization factor $\sqrt{\frac{-g^{+}_{tt}(\theta)}{-g^{-}_{tt}(\theta)}}$ has $\theta$ dependence and does not agree with the Killing vector inside the bubble wall.
Therefore we cannot use the energy conservation with the Komar integral in a consistent way. 

\bibliography{bibs/hoge}
\bibliographystyle{unsrt.bst}

\end{document}